\documentclass[aps,prb,amssymb,amsfonts,eqsecnum,twocolumn,superscriptaddress,showpacs]{revtex4}

\usepackage{bm}
\begin{document}

\title{
Thermoelectric effect in superconducting nanostructures}

\author{V. L. Gurevich}
\affiliation{A. F. Ioffe Physico-Technical Institute, 19021 St. Petersburg, Russia.}
\author{V. I. Kozub}
\affiliation{A. F. Ioffe Physico-Technical Institute, 19021 St. Petersburg, Russia.}
\author{A. L.  Shelankov}
\affiliation{A. F. Ioffe Physico-Technical Institute, 19021 St. Petersburg, Russia.}
\affiliation{Department of Physics, Ume{\aa}  University, SE-901 87 Ume\aa}

\date{19 July 2005}

\begin{abstract}
We study thermoelectric effects in superconducting nanobridges and
demonstrate that the magnitude of these effects can be comparable or
even larger than that for a macroscopic circuit.  It is shown that a
large gradient of the electron temperature can be realistically
created on nanoscale and masking effects of spurious magnetic fields
are minimal in nanostructures. For these reasons nanodevices are
favorable for studying the thermoelectric effect in superconductors.
\end{abstract} 

 \maketitle

\section{Introduction}\label{I}

The discrepancy between the theory and experiment concerning the
thermoelectric phenomena is a long standing problem in physics of
superconductors.  The thermoelectric phenomena in the superconducting
state were first discussed by Ginzburg~\cite{Gi} as early as 1944. In
the presence of a temperature gradient, there appears in a
superconductor a normal current of the form given by
\begin{equation}
 {\bf j}_T=-\eta  \nabla T \,
\label{k4d}
\end{equation}
where $\eta$ is the corresponding transport coefficient. In the bulk
of a homogeneous isotropic superconductor the total electric current
must vanish, and as was pointed out by Ginzburg \cite{Gi}, the normal
current is offset by a supercurrent ${\bf j}_s $ so that the total
current in the bulk
\begin{equation}
 {\bf j}_T+{\bf j}_s=0 \, .
\label{j4d}
\end{equation}
This makes impossible standard studies of the thermoelectric effect in
a homogeneous isotropic superconductor.  Ginzburg considered also
simply-connected anisotropic or inhomogeneous superconductors as
systems where it is possible to observe thermoelectric phenomena by
measuring the magnetic field generated by a temperature gradient.

Theory of the effect was further developed in 1970s~\cite{GGK}.  It
was noted in particular that the offset supercurrent is related to a
difference of the order parameter phases within the simple-connected
superconductor. This phase difference can be measured either in
superconducting interferometer or in the loop formed by different
superconductors where a magnetic flux is generated in the presence of
a temperature difference. It stimulated experimental study of the
effect. Although the first experiment performed by
Zavaritsky~\cite{Zav} is in a rather good agreement with the existing
theory, further experiments (see e.g.~\cite{Garland, vHG}) exhibit
temperature-dependent magnetic fluxes five order of magnitude larger
than is predicted by the theory~\cite{GGK}.  A possibility to observe
large thermoelectric fluxes is discussed in~\cite{my1} and is related
to the phonon drag effect near the contact of the two superconductors
with different values of superconducting gap.  However, the predicted
enhancement factor, the ratio of the Fermi energy and Debye energy, is
not big enough to bridge the gap between the experiment \cite{Garland,
vHG} and the theory.

From the experimental point, the main difficulty is due to the fact
that the thermoelectric effect is small at low temperatures, and one
needs to single it out from various masking effects.  The most obvious
one is related to the temperature dependence of the magnetic field
penetration length~\cite{qwqw,my}. As a result, in the presence of a
background magnetic field, the magnetic field within the
superconductor is temperature dependent. This can mask the genuine
thermoelectric effect. It is important to note that later on it was
shown that the co-existence of a temperature gradient and a
supercurrent leads to variation of the gauge invariant scalar
potential $\phi$ related to an imbalance between the electron-like and
hole-like quasiparticle branches~\cite{Falco1,Schmid,Shel,Clarke}.  In
contrast to the thermoelectric flux, the experimental studies of this
effect were in agreement with the theory~\cite{Clarke}.

The goal of the present paper is to discuss the geometry of experiment
where the thermoelectric effect is particularly large while the
masking effects are significantly suppressed. Therefore one can hope
that it will guarantee unambiguous measurements of the thermoelectric
effect.  With this purpose in mind we will consider thermoelectric
effect in superconducting circuit containing a point contact.

Thermoelectric phenomena in superconducting nanostructures have some
unique specific features that (i) are favorable from the experimental
point of view and (ii) require certain revision of the existing
theory. Experimentally, the advantage is that one is able to create
very large temperature gradients so that the intrinsic thermoelectric
current becomes larger and easier to observe.  We note such a
favorable possibility can be realized only in the systems where the
electrons can be heated as compared to the lattice.  Indeed, a
realistic nanostructures imply a presence of an (insulating) substrate
it is deposited on, and thus large gradients of the lattice
temperature can not be achieved because of the phonon heat conductance
in the substrate. In contrast, the electron temperature, according to
the Wiedemann-Franz law, behaves in the same way as the electrostatic
potential. A detailed discussion demonstrating possibility to obtain
large gradients of electron temperature in metal nanostructures is
given in the Appendices.  Most important, the parasitic effects due to
the trapped magnetic field are much less pronounced in small size
structures.

On the theoretical side, a revision is needed because the earlier
theories have considered bulk samples. Their sizes have been assumed
to be much larger than the characteristic lengths such as the London
penetration length and the length at which the offset supercurrent is
generated. When applied to bulk samples, there is no need to specify
and go into detail of the mechanism by which the normal thermoelectric
current is converted into the offset supercurrent.  This approach is
valid provided the sample size much exceeds the size of the region
where the normal thermoelectric current is converted into the offset
supercurrent. It is well known from the microscopic theory that the
conversion occurs as a result of branch-mixing scattering processes,
where electron-like excitations are scattered to the hole branch of
the excitation spectrum and vice versa.  Microscopic mechanism of the
branch-mixing is known to be inelastic scattering, impurity scattering
in the case of anisotropic superconductors, and Andreev reflection if
inhomogeneity of the order parameter gap is present. If the bulk
scattering is the dominant mechanism, the conversion takes place along
the branch-mixing diffusion length $L_{b}$, $L_{b}= \sqrt{ D
\tau_{b}}$, $D$ and $\tau_{b}$ being the diffusion constant and the
branch-mixing relaxation time, respectively. For a nanostructure of
the size comparable with the branch-mixing length, the standard theory
of the thermoelectric phenomena (that assumes local compensation of
the thermoelectric current) is not applicable. Indeed, in this case
the normal thermoelectric current can be offset also by a normal
diffusion current rather than by a supercurrent \cite{Artem}.

In addition to kinetics, there are important differences in
electrodynamics of superconducting nanostructures. In particular, it
is related to the so-called kinetic inductance ${\cal L}_{\rm k}$
\begin{equation}
{\cal L}_{\rm k} = \frac{L\lambda_{L}^2}{S}.
\label{68d}
\end{equation}
Here $\lambda_{L}$ is the London penetration length, $L$ is the
circuit length while $S$ is the circuit cross-section.  ${\cal
L}_{{\rm k}}$ is inversely proportional to $S$ and may be larger than
the magnetic inductance of the thermoelectric loop for very small
values of $S$.  In this case, the local compensation of the current in
Eq.~(\ref{j4d}) turns out to be energetically unfavorable and the
electrodynamical part of the theory requires a revision too.

In what follows we will develop a theory of thermoelectric effect in
superconducting nanostructures. It will include the above kinetic and
electrodynamical effects.

\section{Charge imbalance distribution}\label{calc1}

Consider two superconducting films (banks) connected by a narrow wire
of the length $L$ and cross-section $S$; the transverse sizes of the
wire are assumed to be much smaller than the London penetration length
$\lambda_{L}$. In this case, the current is distributed homogeneously
across the wire crossection, and the problem is one
dimensional. Denote $x$ the coordinate along the wire and choose the
origin in the middle of the wire. We analyse a diffusive wire and
assume that the temperature varies linearly along the wire between,
its values at the banks being $T_{L}$ and $T_{R}$. Note that the
thermoelectric current is considered a constant equal to $- \eta
\nabla T$ in the wire and zero in the banks. This assumption holds for
3D structures where both the temperature gradient and electric current
density quickly decays within the contact. Naturally, we assume that
the thickness of the wire is much smaller than the thicknesses of the
banks.

First, we briefly overview the well-known physics of branch imbalance
in superconductors. The total electric current density, $\bm{j}=
\bm{j}_{s}+ \bm{j}_{n}$, is a sum of the supercurrent, $\bm{j}_{s}$,
and normal, $\bm{j}_{n}$, components. The supercurrent reads
\begin{equation}
\bm{j}_{s}=
\frac{c^2}{4\pi e \lambda_{L}^2} \bm{p}_{s}
\label{h4d}
\end{equation}
where $\bm{p}_{s}$ is  the superfluity momentum,
\begin{equation}
{\bm p}_{s} = \frac{\hbar}{2}\nabla\chi -
\frac{e}{c}{\bf A} \; , 
\label{3td} \end{equation} 
$\chi $ and $\bf{A}$
being the phase of the order parameter and the vector potential,
respectively.

The normal current,
\begin{equation}
{\bm j}_{n}= {\bm j}_{T}
+ {\bm j}_D \;,
\label{4td}
\end{equation}
is a sum of the thermoelectric current ${\bm j}_{T}= - \eta \nabla T$,
and the diffusion component, $\bm{j}_D=-\sigma\nabla\phi$, related to
the branch imbalance specified by the gauge invariant potential $\phi
$ as ,
\begin{equation}
\phi = \frac{\hbar}{2e} \dot{\chi} +
\varphi \;, \label{i4d}
\end{equation}
$\varphi $ being the scalar potential.  In the vicinity of the
critical temperature, the diffusion current is proportional to the
normal state conductance $\sigma$.

\subsection{Diffusion limit}\label{}

The potentials $\bm{p}_{s}$ and $\phi $ are found from the continuity
equation ${{\rm div}} \bm{j}=0$, and the equation which describes
transformation of the normal current into supercurrent that results in
the following equation for $\phi$ in the wire (see, for instance,
Ref.~\onlinecite{AGGK})

\begin{equation}
\nabla^2 \phi
- \frac{\phi}{L_{b}^2}=0
\label{6td}
\end{equation}
where $\tau_{b}$ is the branch imbalance relaxation time while
$L_{b}=\sqrt{D\tau_b}$ is the branch imbalance relaxation
length~\cite{Tinkham}.  If the banks are made of superconductors with
different gap values, Eq.~(\ref{6td}) requires a boundary condition
\cite{She85} to account for the Andreev reflection at the
interface. The latter plays the role of a surface mechanism of
branch-mixing.

The boundary condition to Eq.~(\ref{6td}) rather generally takes the
form \cite{She85}
\begin{equation}
\left. \frac{1}{\sigma }j_{n} 
\right|_{x=\pm L/2}
= 
\pm\frac{1}{{\cal L}_{b}}
\left.\phi  \rule[-2ex]{0ex}{0ex}\right|_{x=\pm L/2} 
\label{m4d} 
\end{equation}
where $j_{n}$ is the $x-$component of the normal charge current in
Eq.~(\ref{4td}), and ${\cal L}_{b}$ is an effective relaxation length
controlled by the Andreev reflection at the wire-bank interface as
well as the branch-mixing rate in the banks.

Solution to Eq.~(\ref{6td}) with the boundary condition
Eq.~(\ref{m4d}) reads
\[
\phi (x) = \phi _{b} \frac{\sinh \frac{x}{2L_{b}}}
{\sinh \frac{L}{2L_{b}}}
\]
where
\begin{equation}
 \phi_{b}
=
{\displaystyle\frac{1}
{\left(\frac{1}{{\cal L}_{b}}
+
\frac{1}{2L_{b}}\coth \frac{L}{2L_{b}}
\right)}
\frac{j_{T}}{\sigma }}
\; .
\label{qa2}
\end{equation}

The condensate momentum $p_{s}$ and the supercurrent $j_{s}$ is found
from the continuity equation $\text{div} \bm{j}=0$, that is $j(x)=
j_{0}$, where $j_{0}$ is a constant.  From the condition $j_{s}(x) +
j_{n}(x) = j_{0}$ where $j_{n}(x)= j_{T} - \sigma \nabla \phi (x)$,
the distribution of the supercurrent is given by the following
expression
\begin{equation}
j_{s}(x) - j_{0}=- j_{T}\left(
1 - \frac{1}{ \frac{2L_{b}}{{\cal L}_{b}}
+
\coth \frac{L}{2L_{b}}
}
 \;\; \frac{\cosh \frac{x}{2L_{b}}}{\sinh \frac{L}{2L_{b}}}
\right)
\; .
\label{kudd}
\end{equation}
If there is no electrical connection between the banks other than the
wire, the total current must be zero, $j_{0}=0$. Otherwise, the
constant $j_{0}$ is found from electrodynamical considerations
considered in the next section

Thus the temperature difference between the banks leads to a creation
of a potential difference equal to $\Delta \phi =2 \phi_b$ with
$\phi_{b}$ in Eq.~(\ref{qa2}). The potential difference can be
measured, for instance, as described in
Ref.~\onlinecite{Clarke1}. Note that this effect if of the nature
considered by Artemenko and Volkov \cite{Artem}.  However, they
treated a macroscopic circuit with a size $L$ much larger than ${\cal
L}_b$ so that the potential difference was concentrated near the
interface region thus involving only a small part of the total
temperature difference $\Delta T= L \nabla T$.  As a result, their
estimate for "thermoelectric" potential difference is 
\begin{equation}
\Delta \phi \sim \frac{ \eta}{\sigma}\Delta T \frac{{\cal L}_b}{L}
\end{equation}
for a macroscopic wire, the length of which $L$ exceeds the
microscopic scales $ L_{b}$ and ${\cal L}_b$.

In the present paper, we are interested in the opposite limit of a
short wire, $L << L_{b}$.  It follows from Eq.~(\ref{qa2}) that the
potential difference in this limit is
\begin{equation}
\Delta \phi = \frac{j_{T}}{\sigma } 
\frac{1}{\frac{1}{{\cal L}_{b}} + \frac{1}{L}} 
\label{sa2}
\end{equation}

For a short enough wire, $L \lesssim {\cal L}_b$, we obtain
\begin{equation}
\Delta \phi\sim \frac{\eta}{\sigma}\Delta T
\end{equation}
In this case of a short superconducting wire, the thermoelectric
potential difference $\Delta \phi $ is of the order of that in the
normal state.

In a short wire, the supercurrent Eq.~(\ref{kudd}) is homogenous,
\begin{equation}
j_{s} = j_{0} - j_{T}
\frac{L}{ L +{\cal L}_{b}}
\label{ta2}
\end{equation} 
where as before $j_{0}$ is the total electric current through the
wire.

\subsection{Ballistic bridge}\label{}

When we studied the branch imbalance in the previous section, for the
simplicity we have exploited the diffusive approximation.  However,
the largest values of $j_T$ correspond to the largest values of the
mean quasiparticle free path within the wire (leading to larger
$\eta$).  So one expects the largest effect for a ballistic bridge.
In this case one can estimate the quasiparticle thermoelectric current
with the help of a procedure similar to the one used in
Ref.~\onlinecite{Kulik}.  Namely, one has in mind that the
quasiparticle distribution function within the ballistic bridge is
formed by the quasiparticles entering the bridge from the banks. One
also notes that the distribution function is constant along the
quasiparticle trajectory while the quasi-equilibrium distribution
functions of the left and right banks correspond to different
temperatures ($T_L$ and $T_R$, respectively). Thus for the
quasiparticle distribution function, $F_{B}$, within the bridge one
has
\begin{equation}
F_B = \theta\left(v_x \frac{\xi}{\varepsilon}\right)F(T_L) +
\theta\left(-v_x \frac{\xi}{\varepsilon}\right)F(T_R).
\label{qqq}
\end{equation}
Here $v_x$ is the component of the electron velocity along the bridge
direction, and $F(T_{L,R})$ stands for the equilibrium distribution
function corresponding to the temperature $T_{L,R}$.  We have taken
into account that the (group) quasiparticle velocity differs from the
``bare" electron velocity by a factor $\xi/\varepsilon = \xi_{\bm{p}}/
\sqrt{\xi_{\bm{p}}^2 + \Delta^2}$, $\xi_{\bm{p}}$ being the kinetic
energy counted from the Fermi energy, and $\Delta $ being the
superconductor energy gap.  Given the distribution function in
Eq.~(\ref{qqq}), the normal thermoelectric current reads $$
j_{T}=e\sum_{\bm p} v_x F_B \; .$$ As usual in the theory of
thermoelectric phenomena, the contributions of electrons and holes to
the current tend to cancel each other, and the net effect is sensitive
to details of the band structure and to the energy dependence of the
density of states, in particular.  At temperatures $T \gtrsim \Delta
$, the order of magnitude of the thermoelectric current can be
estimated as
\begin{equation} j_T 
\sim ev_F n \frac{(T_L^2- T_R^2)}{\varepsilon_F^2}\;.
\end{equation}
where $v_{F}$ and $\varepsilon_{F}$ are the Fermi velocity and energy,
respectively, and $n$ is the electron density.  For a rough estimate,
assume that the temperature difference is comparable to $T_{c}$.  In
this case,
\begin{equation}
j_T \sim e n v_F \left(\frac{T_{c}}{\varepsilon_{F}}\right)^2 
\label{cur}
\end{equation}
Thus a presence of a temperature drop at the contact between two
superconducting banks leads to formation of the thermoelectric current
through the nanobridge the order of magnitude of which can be
evaluated according to Eq.~(\ref{cur}). The total thermoelectric
current, $I_{T}$, is found by the multiplication of the current
density $j_{T}$ and the bridge cross-section $S$, $I_{T}= j_{T}S.$

\section{Thermoelectric flux}\label{tflux}

As we have discussed above, we study a nanostructure that consists of
a superconducting bridge with a thickness and a width smaller than the
London penetration depth $\lambda_{L}$. The bridge joins two banks
made of the same superconductor (with a critical temperature $T_{c1}$
and a thickness smaller than $\lambda_{L}$). By means of a point-like
heating, using e.g. N-S tunnel junction (see Appendix), the banks are
kept at different temperatures.  We assume that the bridge region
carrying thermoelectric current $I_T$ is short-circuited by
superconducting branch with sizes larger than $\lambda_{L}$ forming a
loop of the linear size $\cal L$. The behaviour of the system is
different for the two limiting cases: a) ${\cal L} > L_b$; b) ${\cal
L} < L_b$.

We start our analysis with the first one, that is the case when the
branch imbalance relaxation length $L_{b}$ is much shorter than the
size of the system. It can be realized in particular if the
near-contact region at least for one of the banks is covered by the
superconductor with a larger gap leading to effective imbalance
relaxation due to Andreev reflections.  If the circuit is
simple-connected the thermoelectric current is compensated by the
supercurrent created due to the Andreev reflection or bulk mechanisms
of the charge imbalance relaxation.  Thus the decay of the normal
thermoelectric current is {\it locally} compensated by supercurrent.

The situation becomes different if the circuit is not
simple-connected, i.e., when another branch (made, for instance, of
the material with a larger $T_c$) closes the loop.  In this case, the
net electric current, built of the normal thermoelectric and
superconducting components, through the bridge may be finite for the
charge current continuity is maintained by the supercurrent $I_{c}$
through the branch closing the loop: $I_{c}$ is actually the electric
current circulating in the loop, and $(I_{c}-I_{T})$ is the
superconducting component of the net current through the bridge.  The
circulating current $I_{c}$ can be readily evaluated minimizing the
total energy $W$ of the system.  The latter is given by the following
expression,
\begin{equation}
W = \frac{1}{2}(I_T - I_c)^2 {\cal L}_{k} + \frac{1}{2}I_c^2{\cal L}
\; .
\label{bum}
\end{equation}
The first term originates from the kinetic energy of superconducting
electrons in the bridge, ${\cal L}_{k} $ being the well-known kinetic
inductance, $${\cal L}_{k} \sim \frac{L\lambda_{L}^2}{S} $$ (where $L$
and $S$, as above, are the bridge length and cross-section,
respectively). The second term in Eq.~(\ref{bum}) is the energy of
magnetic field created by the circulating current $I_{c}$, and $\cal
L$ is the inductance of the loop, which is close to the geometrical
inductance of the macroscopic branch.  Minimizing $W$ with respect to
$I_c$, one obtains
\begin{equation}
I_c =
I_T \frac{{\cal L}_k }{{\cal L}_k + {\cal L}}
\end{equation}
and, thus the thermoelectric magnetic flux is
\begin{equation}
\Phi_T = I_T\frac{{\cal L}_k {\cal L}}{{\cal L}_k + {\cal L}}.
\label{qq}
\end{equation}
The flux $\Phi_{T}$ is controlled by the smaller of the inductances in
question.

Note that if ${\cal L}_k \ll \cal L,\;$ $\Phi_T$ does not depend on
$\cal L$ and is estimated as
\begin{equation}
\label{1}
\Phi_T  = I_{T} {\cal L}_{k}\sim I_T \frac{L\lambda_{L}^2}{S} \; .
\label{bim}
\end{equation}
In the dirty limit, the penetration depth $\lambda_{L}$ is related to
its value in thebulk pure material as $$\lambda_{L}^2 = \lambda_0^2
(\xi_0/l_e)$$ where $\xi_0 \sim v_{F}/\Delta$ is the coherence length,
and $l_e$ is the electron elastic mean free path.  As it can be seen,
this result coincides with the predictions of the papers
\cite{GGK,Garland} assuming that the thermoelectric current is almost
completely compensated by the supercurrent.

However, the situation is qualitatively different if ${\cal L}_k \gg
\cal L$, a condition which can be realistically met for a nanoscale
bridge.  Indeed, assuming $L \sim \sqrt{S}$ and $L \sim l_{e}\sim
10^{-6}cm$, for $\lambda_0 \sim 10^{-5}$cm, $\xi_0 \sim 10^{-4}$cm,
one obtains ${\cal L}_k \sim 10^{-2}$cm. This means that the kinetic
inductance ${\cal L}_{k}$ may be comparable to the magnetic geometric
inductance ${\cal L}$ even for a relatively large, nearly macroscopic
loop.  In this case, the normal thermoelectric current generated by
the bridge is non-locally short-circuited by the supercurrent through
the macroscopic branch than being offset locally by the supercurrent.
In this limit, one has from Eq.~(\ref{qq}):
\begin{equation}\label{2}
\Phi_T = I_T {\cal L}.
\end{equation}
Despite the absence of the current cancellation within the bridge, the
flux through the loop is ${\cal L}_{k}/{\cal L}$ times {\it smaller}
than it has been predicted in earlier papers \cite{GGK,Garland}. At
the same time, the magnetic field within the structure practically
coincides with its value for a normal metal
structure. Correspondingly, {\it it can be much larger than for the
thermoelectric effect in macroscopic circuits.} Indeed, assuming that
the inductance $\cal L$ is of the order of the linear size of the
circuit, our estimates for the magnetic field from Eqs.~(\ref{1}), and
(\ref{2}) are
\begin{equation}
H_T \sim I_T \frac{{\cal L}_k}{{\cal L}^2}
\quad , \quad {\cal L}_{k}\ll {\cal L} 
\end{equation}
and
\begin{equation}
H_T \sim \frac{I_T}{\cal L}
\quad , \quad {\cal L}_{k}\gg {\cal L} 
\end{equation}
Thus the "thermoelectric" magnetic field is the larger the smaller is
$\cal L$ and is much larger for the regime ${\cal L}_k > {\cal L}$
than for a ``macroscopic'' considered earlier in \cite{GGK,Garland}.
We believe that this factor significantly suppresses a possible role
of masking effects.

In the limiting case (b), when the charge imbalance length $L_{b}$ is
much shorter than the size of the system (${\cal L} < L_b)$, the
quasiparticle thermoelectric current is not converted into a
supercurrent but short-circuited by the normal current through the
closing branch (as it occurs in normal metal thermoelectric
circuits). The normal charge current in the loop generates a magnetic
flux which in turn generates a circulating supercurrent $I_{c}$ in the
direction opposite to the normal current.  In this case, the energy
$W=I_c^2 {\cal L}_k/2 + (I_T - I_c)^2 {\cal L}/2$ is built of the
supercurrent kinetic energy $I_c^2 {\cal L}_k/2$ and the magnetic
energy $(I_T - I_c)^2 {\cal L}/2$.  Minimizing $W$, $$ I_c = I_T
\frac{\cal L}{{\cal L} + {\cal L}_k}\; , $$ and the total
thermoelectric flux, $\Phi_T = (I_T - I_c){\cal L}$, is again given by
Eq.~(\ref{qq}).  Therefore, the thermoelectric flux $\Phi_{T}$ is
completely controlled by the normal component if ${\cal L}_k > {\cal
L}$.

Let us estimate the largest possible values of $\Phi_T$ which can be
realized for large $\cal L$. We have
\begin{equation}
\Phi_T \sim I_T \frac{L\lambda^2}{S}
\end{equation}
where $L$ and $S$ are the bridge length and cross-section
respectively.

Correspondingly,
\begin{equation}
\Phi_T\sim  e n v_F
\left(\frac{T}{\mu}\right)^2\frac{L\lambda_0^2\xi_0}{l_e}.
\end{equation}
In what follows we will assume that all the sizes of the
bridge are of the same order while one should also put
$l_e \sim L$.  Assuming $T/\mu \sim 10^{-4}$ ($T \sim 1 $K), $\lambda^2
\sim 10^{-10}\,$cm$^2$, $\xi_0 \sim 10^{-4}$\,cm one has $\Phi_T \sim
10^{-3}\Phi_0$.

For smaller $\cal L$ the magnetic fluxes are smaller than the above
estimate but the magnetic fields are higher.

\section{Conclusion}\label{C}

In this paper we have analyzed the thermoelectric effects in
superconducting nanostructures.  When the size of a thermoelectric
circuit is less than the branch imbalance length, the very picture of
the thermoelectric effects becomes different from that considered
earlier for macroscopic systems: rather than being offset locally by
the supercurrent, the quasiparticle thermoelectric current is
short-circuited nonlocally, by the diffusion current in the branch
closing the circuit, similar to the picture of the effect in normal
metal thermoelectric loops.  The thermoelectric effects in
superconducting nanostructures may be comparable with that in systems
of macroscopic size systems.  At the same time, the masking effects
inherent for macroscopic superconductors can be eliminated so that
nanoscale structures are promising for studying the thermoelectric
effects in superconductors.

\acknowledgements The authors acknowledge support for this work by the
grant of Swedish Royal Academy.  V. L. G. and V. I. K. also
acknowledge partial support for this work by the Russian National
Foundation for Basic Research, grant No~03-02-17638.

\appendix

\section{Temperature distribution}\label{T}

The purpose of this section is to discuss the conditions when one can
ascribe different temperatures to elections in two banks connected by
a short bridge.

In recent years it has been demonstrated that the electronic
temperature of a metal film may substantially differ from the lattice
temperature of the dielectric substrate.  For quasi-2D metallic
nanostructures at low temperatures, there are two factors that are
favorable for such a possibility.  First, small electron-phonon
collision rates prevent effective transfer of heat to the phonon
system of the substrate. Second, the phonon heat conductivity of the
substrate at small spatial scales turns out to be smaller than the
electron heat conductivity within the films since the  phonon
mean free path is limited by the spatial inhomogeneity. Using the
Wiedemann~-~Franz law, one estimates the electron heat current within
the metal layer of a length $L$ and cross-section $S$ as
\begin{equation}
Q_{el} \sim \frac{\Delta T}{L}SD_e{\tilde n}
\label{zce}
\end{equation}
where $\Delta T$ is the temperature difference, $D_e$ is electron
diffusivity, and $\tilde n \sim (T/\varepsilon_F)n$ is the
concentration of quasiparticles participating in the heat transfer
(where $n$ is the total electron concentration while $\varepsilon_F$
is the Fermi energy).  At the same time, the heat flux from the film
to the substrate can be estimated as
\begin{equation}
Q_{\hbox{sub}} \sim \frac{SL{\tilde n}\Delta T}{\tau_{{\rm e-ph}}}
\label{2ce}
\end{equation}
where $\tau_{{\rm e-ph}}$ is electron-phonon relaxation time.  From
Eqs.~(\ref{zce}), and (\ref{2ce}), one sees that $Q_{\rm el} > Q_{{\rm
sub}}$ provided $L^2 < l_e v_F \tau_{{\rm e-ph}}$.

It is also instructive to compare the electronic heat flux $Q_{el}$
with the heat flux $Q_{ph}$ supported by phonons in the substrate and
``shunting'' the electron flux.  One easily obtains that $Q_{el} >
Q_{ph}$ provided
\begin{equation}
\frac{L}{d} \frac{s \min (w, l_{ph})}{v_F
l_e}\left(\frac{T}{T_D}\right)^3
\frac{\varepsilon_F}{T} < 1
\end{equation}
where $T_D$ is the Debye temperature of the substrate, $w$ is the
width of the metal layer, $d$ is the layer thickness, $s$ is the sound
velocity while $l_e$ and $l_{ph}$ are the mean free paths of electrons
within the layer and phonons within the substrate, respectively. Since
the electron heat conductivity dominates provided any of the
aforementioned conditions holds, one concludes that at low
temperatures the electron temperature is mainly controlled by electron
heat conductivity of the metal structure.

Consider now a point ballistic bridge connecting to metal banks with
different electronic temperatures. It follows from the above
considerations that the temperature drop is concentrated mainly within
the contact region.  Indeed, for 3D geometry and a diffusive transport
in the bulk, the temperature distribution in the banks near the bridge
follows the same law as an electric potential distribution, that is
the temperature drop is concentrated in the bridge.  If the whole
structure is made of a metal film of the same thickness and with the
diffusive electron transport this statement holds only with a
logarithmic accuracy because of the 2D character of electron
diffusion. However if the thickness of the bridge region is much
smaller than the thicknesses of the banks (that is if the
configuration is a 3D-like one) the temperature drop is again
completely restricted by the contact region. The same holds provided
the electron transport within the contact and near-contact regions is
ballistic. Indeed, it follows from the fact that under the
Wiedemann-Franz law the temperature profile is similar to the electric
potential profile while in 2D ballistic structures the potential drop
is concentrated in the contact region.

It is expected that very large values of $\Delta T$ can be realized in
the point contact geometry.  Indeed, one can apply for the heat flux
the same arguments as for electric current through the point contact
\cite{Kulik}, namely, that the relaxation processes for the electrons
take place within the bulk of the sample at distances $(\sim D_e
\tau_{ee})^{1/2}$. Thus enormous values of temperature gradient and
heat flux density do not lead to destruction of the bridge.

\subsection{Electron heating}\label{}

Let us consider the important practical question concerning the
generation of the temperature gradient across the bridge.  We have
assumed above that the excitations within the one of the banks are
heated as compared to the excitations in another one. Since we deal
with superconductors, it excludes the Joule heating. On the other
hand, microwave heating implies relatively large areas. In our
opinion, the best way is to heat electrons on one on the banks using a
tunnel S-I-N junction.  The junction is formed by a normal film of
area $S_{2}$ put on the top of the superconducting bank (with a thin
insulating layer).  When the bias $eV $ across the S-I-N junction is
much larger than the superconductor energy gap, high-energy electrons
tunnelling from N layer will relax mainly due to creation of
electron-hole pairs within the superconducting layer.  To have the
electron temperature formed, one should have $$ S_{2} > v_F l_e
\tau_{ee}, $$ $ \tau_{ee}$ being the electron-electron scattering
time.

Now let us compare the thermal current from the heated superconducting
layer to the substrate and the thermal current through the point
contact to the ``cold'' bank. Assuming that the thickness of the
superconducting layer forming the tunnel junction and the layer
forming the point contact are the same, one finds that the thermal
current through the contact dominates provided
\begin{equation}
S_2 L/w < l_e v_F \tau_{\hbox{e-ph}}
\end{equation}
where $L$ and $w $ are the point contact length and width,
respectively. If $L \sim w $ one concludes, that this condition can
hold since at low enough temperatures $\tau_{e_ph} > \tau_{ee}$.
Correspondingly, in this case only a region with the area $S_2$ (under
the tunnel junction) is heated with respect to the rest of the device,
the heat leak being due to thermal current through the point
contact. Certainly, one should also assume that the area of the
superconducting layer in the ``cold'' bank is large enough to ensure
efficient heat withdrawal to its substrate.  In this case one easily
obtains
\begin{equation}
\Delta T \sim IV \frac{L\varepsilon_F}{wdD_enT}
\end{equation}
where $I$ is the current through the tunnel junction.

The main conclusion following from the considerations given above is
that it is possible to have ``point-like'' electron heating restricted
by the area $\sim v_F l_e \tau_{ee}$. Its linear dimensions for
realistic estimates can be as small as 3 $\mu$m. Correspondingly, if
the inductance loop has macroscopic size this local heating (and
corresponding local variation of the penetration length) is not
expected to affect the temperature-dependent (or rather $V$-dependent)
flux through the loop.


\begin{thebibliography}{99}

\bibitem{Gi} V. L.  Ginzburg, Zh.
 Eksp.  Teor. Fiz. {\bf14}, 177 (1944).

\bibitem{GGK}Yu. M. Galperin, V. L. Gurevich, V. I. Kozub, Sov. Phys. -
JETP Letters, {\bf 17}, 476 (1973); Yu. M. Galperin, V. L. Gurevich,
V. I. Kozub, Sov. Phys. - JETP, {\bf 65}, 1045 (1974)

\bibitem{Zav} N. V. Zavaritskii, Pis'ma Zh. Eksp. Teor. Fiz. {\bf 20},
 223 (1974) [JETP Lett. {\bf 20}, 97 (1974)].

\bibitem{Garland} D. G. Van Harlingen, J. C. Garland, Solid State
 Commun. {\bf 25}, 419 (1978).

\bibitem{vHG} D. J. Van Harlingen, D. F. Heidel, and J. C. Garland,
 \prb {\bf 21}, 1842 (1980).

\bibitem{my1} V. I. Kozub, Zh. Eksp. Teor. Fiz. {\bf 88}, 1847 (1985)
 [Sov. Phys. JETP, {\bf 61}, 1095 (1985)]

\bibitem{qwqw} C. M. Pegrum, A. M. Guenault, Phys. Letters, {\bf 59
A}, 393 (1976)

\bibitem{my} V. I. Kozub, Sov. Phys. - JETP, {\bf 47}, 178 (1978).

\bibitem{Falco1} C. M. Falco, Phys. Rev. Lett. {\bf 39}, 660 (1977).

\bibitem{Schmid} A. Schmid, G. Sch\"on, Phys. Rev. Lett. {\bf 43}.
 793 (1979).

\bibitem{Shel} A. L. Shelankov,
 Zh. Eksp. Teor. Fiz. {\bf 78}, 2359 (1980) [Sov. Phys. - JETP,
 {\bf 51}, 1186 (1980)].

\bibitem{Clarke} J. Clarke, B. R. Fjordboge, P. E. Lindelof,
 Phys. Rev. Lett. {\bf 43}, 642 (1979).

\bibitem{Artem} N.Artemenko, A.F.Volkov Sov. Phys.-JETP, {\bf 43},
548 (1976)

\bibitem{AGGK}A. G. Aronov, Yu. M. Galperin, V. L. Gurevich and
V. I. Kozub, in : Nonequilibrium superconductivity,
ed. by V. M. Agranovich, A. A. Maradudin, North Holland, 1976.

\bibitem{Tinkham}J. Clarke and M. Tinkham, \prl {\bf44}, 106 (1980).

\bibitem{She85} A. L. Shelankov,
Sov. Phys. Solid State {\bf 27}, 965 (1985).

\bibitem{Clarke1} J. Clarke, in : Nonequilibrium superconductivity,
ed. by V. M. Agranovich, A. A. Maradudin, North Holland, 1976.

\bibitem{Kulik} I. O. Kulik, A. N. Omel'yanchuk, R. I. Shekhter,
Solid State  Comm. {\bf 23}, 301 (1977)


\end{thebibliography}
\end{document}